\documentclass[12pt]{iopart}
\usepackage{graphicx}
\usepackage{amssymb}

\def\Journal#1#2#3#4{{#1} {\bf #2}, (#3) #4}

\def\etal{{\it et al.}}
\def\AA{\em A.\& A.}

\def\APJ{\em ApJ.}

\def\IET{\em IEEE Trans. Comm.}

\def\MRA{\em MNRAS}

\def\NPB{{\em Nucl. Phys.} B}

\def\PLB{{\em Phys. Lett.}  B}

\def\PRD{{\em Phys. Rev.} D}

\def\filepath{./} 

\def\be{\begin{equation}}
\def\ee{\end{equation}}
\def\bea{\begin{eqnarray}}
\def\eea{\end{eqnarray}}

\begin{document}
\title[Nonparametric determination of redshift evolution]{Nonparametric 
determination of redshift evolution index of Dark Energy}
\author{Houri Ziaeepour}
\address {Mullard Space Science Laboratory, Holmbury St. Mary, Dorking, 
Surrey, RH5 6NT, UK.}
\ead{hz@mssl.ucl.ac.uk}


\begin {abstract}
We propose a nonparametric method to determine the sign of $\gamma$ - the 
redshift evolution index of dark energy. This is important for distinguishing 
between positive energy models, a cosmological constant, and what is 
generally called ghost models. Our method is based on geometrical properties 
and is more tolerant to uncertainties of other cosmological parameters than 
fitting methods in what concerns the sign of $\gamma$. The same 
parameterization can also be used for determining $\gamma$ and its redshift 
dependence by fitting. We apply this method to SNLS supernovae and to gold 
sample of re-analyzed supernovae data from Riess \etal 2004. Both datasets 
show strong indication of a negative $\gamma$. If this result is confirmed 
by more extended and precise data, many of dark energy models, including 
simple cosmological constant, standard quintessence models without interaction 
between quintessence scalar field(s) and matter, and scaling models are ruled 
out. 
\end {abstract}


\maketitle

Recent observations of Supernovae (SNe), Cosmic Microwave Background (CMB), 
and Large Scale Structures (LSS) indicate that the dominant content of the 
Universe is a mysterious energy with an equation of state very close to 
Einstein cosmological constant. The equation of state is defined by $w$, 
the ratio of pressure $p$ to density $\rho$, $w = P/\rho$. For a cosmological 
constant $w = -1$. The observed mean value of $w$ for dark energy is very 
close to $-1$. Some of the most recent estimations of $w$ are: From 
combination of 3-year WMAP and SuperNova Legacy Survey (SNLS), 
$w = -0.97^{+0.07}_{0.09}$\cite{wmapeyear}; from combination of 3-year WMAP, 
large scale structure and supernova data, 
$w = -1.06^{+0.016}_{-0.009}$\cite{wmapeyear}; from combination of CMAGIC 
supernovae analysis and baryon acoustic peak in SDSS galaxy clustering 
statistics at $z = 0.35$, $w = -1.21^{+0.15}_{-0.12}$\cite{cmagicsn}; 
and finally from baryon acoustic peak alone $w = -0.8 \pm 0.18$. It is evident 
that with inclusion of one or two sigma uncertainties to measured mean values, 
the range of possible values for $w$ runs across the critical value of 
$-1$. Moreover, in all these measurements the value of $w$ depends on 
other cosmological parameters and their uncertainties in a complex way.
Reconstruction methods for determining cosmological parameters from 
observations - see \cite{reconsrev} and references therein for a review of 
methods - usually use fitting of continuous parameters on the data and 
determine a range for $w$. On the other hand, the sign of $w+1$ is more 
crucial for distinguishing between various dark energy models than 
its absolute value. For instance, if $w+1 < 0$, quintessence models with 
conventional kinetic energy and potential are ruled out because in these 
models this quantity is always positive. Decay of dark matter to dark energy 
or an interaction between these components can lead to an effective 
$w+1 < 0$ without violating null energy condition\cite {quinint}. Redshift 
dependence of $w$ has been invoked in many of recent works. Although 
the quality of available data does not yet permit to obtain a precise 
measurement of variability of $w$, recent attempts show that at low redshifts 
this can not be very large~\cite{wvar}, see also the results of present work 
below.

Without loss generality the density of the Universe at redshift $z$ can be 
written in the following form:
\be
\frac {\rho (z)}{\rho_0} = \Omega_m (1+z)^3 + \Omega_h (1+z)^4 + 
\Omega_{de} (1+z)^{3\gamma} 
\label{density}
\ee
where $\rho (z)$ and $\rho_0$ are total density at redshift $z$ and in 
local Universe, respectively; $\Omega_m$, $\Omega_h$, and $\Omega_{de}$ 
are respectively cold and hot matter, and dark energy fraction in the 
total density at $z=0$. We consider a flat universe in accordance with 
recent observations\cite{wmapeyear}. At low redshifts, the contribution 
of the CMB to the total mass of the Universe is negligible. The contribution 
of neutrinos is ${\Omega}_{\nu} h^2 = \sum m_{\nu} / 92.8~eV$, 
$h \equiv H_0 / 100 km~Mpc^{-1}sec^{-1}$, where $H_0$ is present Hubble 
constant. The upper limit on the sum of the mass of neutrinos from 
3-year WMAP is $\sum m_{\nu} < 0.62$ ($95\%$ confidence 
level)\cite{omeganu}. Even if one of the neutrinos has very small mass and 
behaves as a warm dark matter, for $z < 1$ neutrino contribution to the 
total mass is $\lesssim 4\%$, smaller than the uncertainty on dark matter 
contribution. Thus, the approximation $\Omega_m + \Omega_{de} \approx 1$ is 
justified. 

Here we propose a nonparametric method specially suitable for 
estimating the sign of $\gamma$. When the quality of data is adequate, the 
quantity $A(z)$ defined in (\ref{densderiv}) can also be used to fit the data 
and to measure the value of $\gamma$. The expression {\it nonparametric} here 
is borrowed from signal processing literature where it means testing a null 
hypothesis against an alternative hypothesis by using a discrete condition 
such as jump, sign changing, etc., in contrast to constraining a continuous 
parameter (see e.g. \cite{nonparam}). We show that geometrical 
properties of $A(z)$ are related to the sign of $\gamma$ and we can detect 
sign without fitting a continuous parameter.

Assuming a constant $w$
\be
\gamma = w+1  \label {gammaconst}
\ee
and it can be easily shown that in this case:
\be
{\mathcal A}(z) \equiv \frac{1}{3 (1+z)^2 \rho_0} \frac {d\rho}{dz} - 
\Omega_m = \gamma \Omega_{de}(1+z)^{3 (\gamma - 1)} \label {densderiv}
\ee
If the equation of state of dark energy varies with redshift, we can assume 
that $\gamma$ in (\ref{density}) depends on redshift. In this definition 
$\gamma$ is related to the equation of state of dark energy according to:
\be
\gamma (z) = \frac{1}{\ln (1+z)} \int_0^z dz'\frac{1 + w(z')}{1+z'} 
\label {gammaz}
\ee
and:
\be
A(z) = \Omega_{de} \biggl (\gamma (z) + (1+z) \ln (1+z) \frac {d\gamma}{dz}
\biggr) (1+z)^{3 (\gamma -1)} \label {densdervar}
\ee
When $w$ is constant (\ref{gammaz}) and (\ref{densdervar}) approach 
respectively to (\ref{gammaconst}) and (\ref{densderiv}). If $w (z)$ is a 
slowly varying function of redshift, we can parameterize it as:
\be 
w (z) = w_0 + w_1 z \label {wz}
\ee 
where $w_0$ and $w_1$ are constant, and $|w_1 / w_0| < 1$. This leads to:
\be
\gamma (z) = 1 + w_0 + w_1 (z /\ln (1+z)-1) \label {gammazw}
\ee
By expanding the logarithm term at low redshifts it is straightforward to see 
that at lowest order $\gamma \approx 1 + w_0 + w_1 z/2$. Therefore 
$\gamma (z)$ varies even more slowly than $w$. Using (\ref{gammazw}) and  
(\ref{densdervar}), we can find the explicit expression of $A(z)$ as a 
function of constant parameters:
\bea
A(z) & = &\Omega_{de} \biggl (\gamma (z) + \frac {w_1 [(1+z) \ln (1+z) - z]}
{\ln (1+z)}\biggr) (1+z)^{3 (\gamma -1)} \nonumber \\ & \approx & \Omega_{de} 
\biggl (\gamma (z) + w_1 (\frac {z^2}{2} + \ldots) \biggr) 
(1+z)^{3 (\gamma -1)} \label{aw}
\eea
This shows that at low redshifts the contribution of derivative term is 
quadratic in $z$ and negligible in comparison with $\gamma (z)$. 
Therefore, at low redshifts the sign of $A(z)$ follows the sign of 
$\gamma (z)$ even for a redshift-dependent equation of state. On the other 
hand, for $z <1$ this parameterization is equivalent to models IV and IV of 
Serra \etal~\cite{wvar}, which are barely constrained and are 
consistent with zero or small $w_1$. This confirms that the assumption of a 
constant or slowly varying $\gamma$ at low redshifts in (\ref{densderiv}) is 
a valid approximation for our Universe. Therefore in this work from now on we 
consider only this case, unless the redshift dependence of $\gamma$ is 
explicitly mentioned.

Similar expressions can be obtained for non-standard cosmologies such as 
DGP\cite{dgp} model and other string/brane inspired 
cosmologies~\cite{branecosmos}. It is also possible to find an expression 
similar to (\ref{densderiv}) for non-flat FLRW models and without neglecting 
hot matter. The left hand side would however depend on $\Omega_k$ and 
$\Omega_h$, and therefore would be more complex. Nonetheless, when their 
contribution at low redshifts are much smaller than cold matter and dark 
energy, the general behaviour of ${\mathcal A}(z)$ will be the same as 
approximate case studied here.

In summary, the right hand side of (\ref{densderiv}) has the same sign as 
$\gamma$ if $|\gamma (z)| \gg (1+z) \ln (1+z) |\gamma'(z)|$. Moreover, the 
sign of $A (z)$ derivative is 
opposite to the sign of $\gamma$ because due to the smallness of observed 
$\gamma$, the term $\gamma -1$ is negative (if $\gamma$ depends on $z$ this 
is true only for low redshifts). This means that ${\mathcal A}(z)$ is a 
concave or convex function of redshift, respectively for positive or negative 
$\gamma$. In the case of a cosmological constant 
${\mathcal A}(z) = 0$ for all redshifts. This second feature of expression 
(\ref{densderiv}) is interesting because if $\Omega_m$ is not correctly 
estimated, ${\mathcal A}(z)$ will be shifted by a constant, but its 
geometrical properties will be preserved.

The left hand side of expression (\ref{densderiv}) can be directly estimated 
from observations. More specifically, $\Omega_m$ is determined from 
conjunction of CMB, LSS, and supernova type Ia data, and at present it 
is believed to be known with a precision of $\sim 5\%$. At low redshifts, 
the derivative of the density is best estimated from SN type Ia observations. 
In the case of FLRW cosmologies, the density and its derivative can be 
related to luminosity distance $D_l$, and its first and second derivatives:
\bea
&& {\mathcal B}(z) \equiv \frac{1}{3 (1+z)^2 \rho_0} \frac {d\rho}{dz} = 
\frac{\frac{2}{1+z}(\frac{dD_l}{dz} - \frac {D_l}{1+z}) - 
\frac{d^2D_l}{dz^2}}{\frac{3}{2} (\frac{dD_l}{dz} - \frac {D_l}{1+z})^3}
\label {rhoderdl} \\
&& D_l = (1+z) H_0 \int_0^z \frac {dz}{H (z)} \quad , \quad 
H^2 (z) = \frac{8\pi G}{3} \rho (z) \label {dlh}
\eea
It is remarkable that the right hand side of (\ref{rhoderdl}) depends only 
on one cosmological parameter, $H_0$. Moreover, similar to an uncertainty 
on $\Omega_m$, $H_0$ uncertainty rescales ${\mathcal B}(z)$ similarly at all 
redshifts, but does not modify geometrical properties ${\mathcal B}(z)$ and 
${\mathcal A}(z)$. $D_l$ can be directly obtained from observed luminosity of 
standard candles such as supernovae type Ia. In the case of LSS observations 
where the measured quantity is the evolution of $\rho$ with redshift, 
${\mathcal B}(z)$ is measured directly up to an overall scaling by 
$\rho_0$. This does not change the geometrical properties of ${\mathcal B}(z)$ 
and ${\mathcal A}(z)$ either. The scaling by a positive constant does not 
flip a convex curve to concave or vis-versa. In summary, uncertainties of 
$\Omega_m$ and $H_0$ do not affect the detection of the sign of $\gamma$ 
through geometrical properties of ${\mathcal A}(z)$. This is quite different 
from fitting methods. They are sensitive to all numerical parameters $H_0$, 
$\Omega_m$, $\Omega_{de}$, and $w$ in a complex way, usually through a 
non-linear equation such as chi-square or likelihood. This makes the 
assessment of the influence of uncertainties on parameter estimation, and 
specifically on the determination of the sign of $\gamma$ very difficult. 

When this method is applied to standard candle data such as supernovae type 
Ia, one has to calculate first and second derivatives of $D_l$. Numerical 
calculation of derivatives is not however trivial. 
To have a stable and enough precise result, not only the data must have high 
resolution and small scatter, but also it is necessary to smooth it. Therfore, 
to test the stability of numerical calculations and the method in general, we 
also apply it to simulated data. The details of numerical methods is discussed 
in the Appendix. 

\begin{figure}[h]
\begin{center}
\begin{tabular}{cc}
{\bf a} \includegraphics[width=6cm]{\filepath/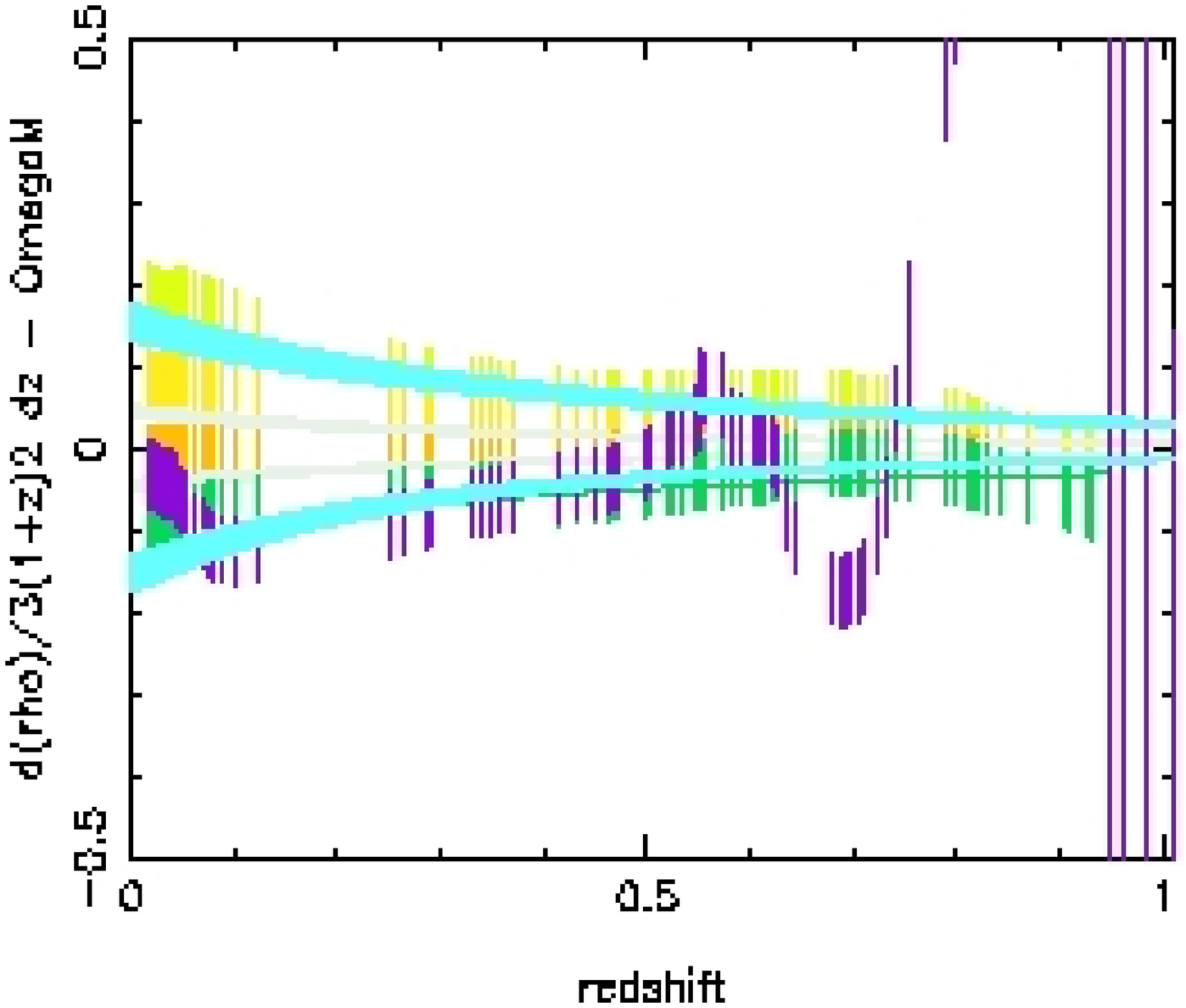} &
{\bf b} \includegraphics[width=6cm]{\filepath/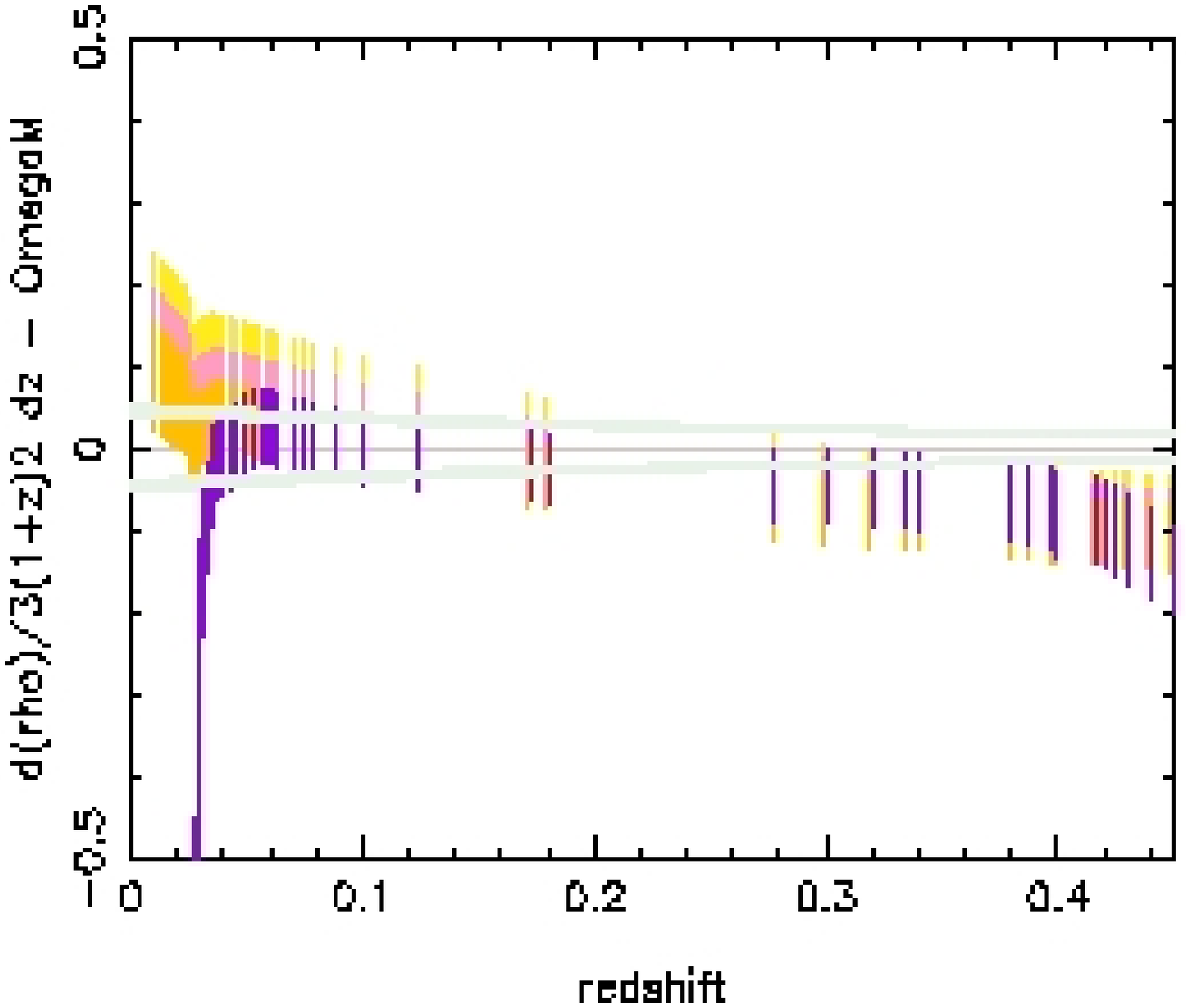}
\end{tabular}
\caption{${\mathcal A}(z)$ from 117 SNe of the SNLS 
({\bf a}-purple curve), and from 88 SNe with $z < 0.45$ recompiled and 
re-analyzed by Riess \etal 2004\cite{sncp} ({\bf b}-purple curve). Error 
bars are 1-sigma uncertainty. In ({\bf a}), green, orange, yellow, and 
light green curves are the reconstruction of ${\mathcal A}(z)$ from 
simulations for $\gamma = -0.2, -0.06, 0.6, 0.2$, respectively. The 
probability of null hypothesis ($\gamma = 0$) is $\bar{P} = 0.27$, therefore 
the probability of $\gamma < 0$, $1-\bar{P} = 0.73$. Light grey and cyan 
curves are theoretical calculation including the uncertainty on $\Omega_{de}$, 
respectively for $\gamma = \pm 0.06, \pm 0.2$. In ({\bf b}), the pink curve 
is simulated $\gamma = 0$ model. For this dataset $1 - \bar{P} = 0.75$. The 
dark grey straight line is the 
theoretical expectation for ${\mathcal A}(z)$ when dark energy is a 
cosmological constant. For all models $H_0 = 73~km~Mpc^{-1}~sec^{-1}$ and 
$\Omega_{de} = 0.77$. 
\label{fig:sncp}}
\end{center}
\end{figure}

Examples of reconstruction of simulated data are shown in 
Fig.\ref{fig:sncp}. It is evident that despite deformation of the 
reconstructed curves due to numerical errors and uncertainties, specially 
close to lower and upper redshift limits, the difference between convexity of 
curves for positive and negative $\gamma$ is mostly preserved and can be used, 
either by visual inspection or by using a slope detection algorithm, to 
find the sign of $\gamma$. The simulated data is however much more uniform - 
has less scatter - than presently available data, see Fig.\ref{fig:sncp}. 
Therefore the reconstruction from real data is more prone to artifacts than 
simulated ones. Fig.\ref{fig:sncp} shows that although at mid-range redshifts 
the data follow a curve similar to what is expected from FLRW cosmologies, 
deviations appear close to boundaries and at high redshifts where the quality 
of data is worse. We attribute these to the artifacts of reconstructions.

Visual inspections or slope detection lack a quantitative estimation of 
uncertainties of the measured sign for $\gamma$. In signal processing usually 
binomial estimation of the probability or optimization of 
detection\cite{signopt} is used to assess uncertainties. However, in most 
practically interesting contexts in signal processing, the signal is constant 
and uncertainties are due to noisy. In the cosmological sign detection 
discussed here the observable ${\mathcal A}(z)$ is both noisy and varies with 
redshift. Therefore, binomial probability and similar methods are not 
applicable. For this reason we try another strategy, specially suitable for 
this cosmological sign detection task. 

The null hypothesis for dark energy is $\gamma = 0$. Assuming a Gaussian 
distribution for the uncertainty of reconstructed ${\mathcal A}(z)$ from data 
and from simulated data with $\gamma = 0$, for 
each data-point we calculate the probability that the data-point belongs to 
the null hypothesis. To include the uncertainty of data, we integrate the 
uncertainty distribution 1-sigma around the mean value:
\be
P_i = \frac{1}{\sqrt{2\pi (\sigma_{0i}^2 + \sigma_{i}^2)}} \int_{{\mathcal A}_i-\sigma_i}^{{\mathcal A}_i+\sigma_i} dx e^{-\frac {(x - {\mathcal A}_{0i})^2}{2 (\sigma_{0i}^2 + \sigma_{i}^2)}} 
\label{probzero}
\ee
where ${\mathcal A}_i$ and $\sigma_i$ belong to the $i^{th}$ data-point, and 
${\mathcal A}^{0i}$ and $\sigma_{0i}$ belong to simulated null hypothesis 
model at the same redshift. Averaging over $P_i$ gives $\bar {P}$, an overall 
probability that the dataset corresponds to the null hypothesis. As 
$\gamma = 0$ is the limit case for $\gamma > 0$, $\bar {P}$ is also the 
maximum probability of $\gamma > 0$.

\begin{figure}[h]
\begin{center}
\begin{tabular}{cc}
{\bf a} \includegraphics[width=6cm]{\filepath/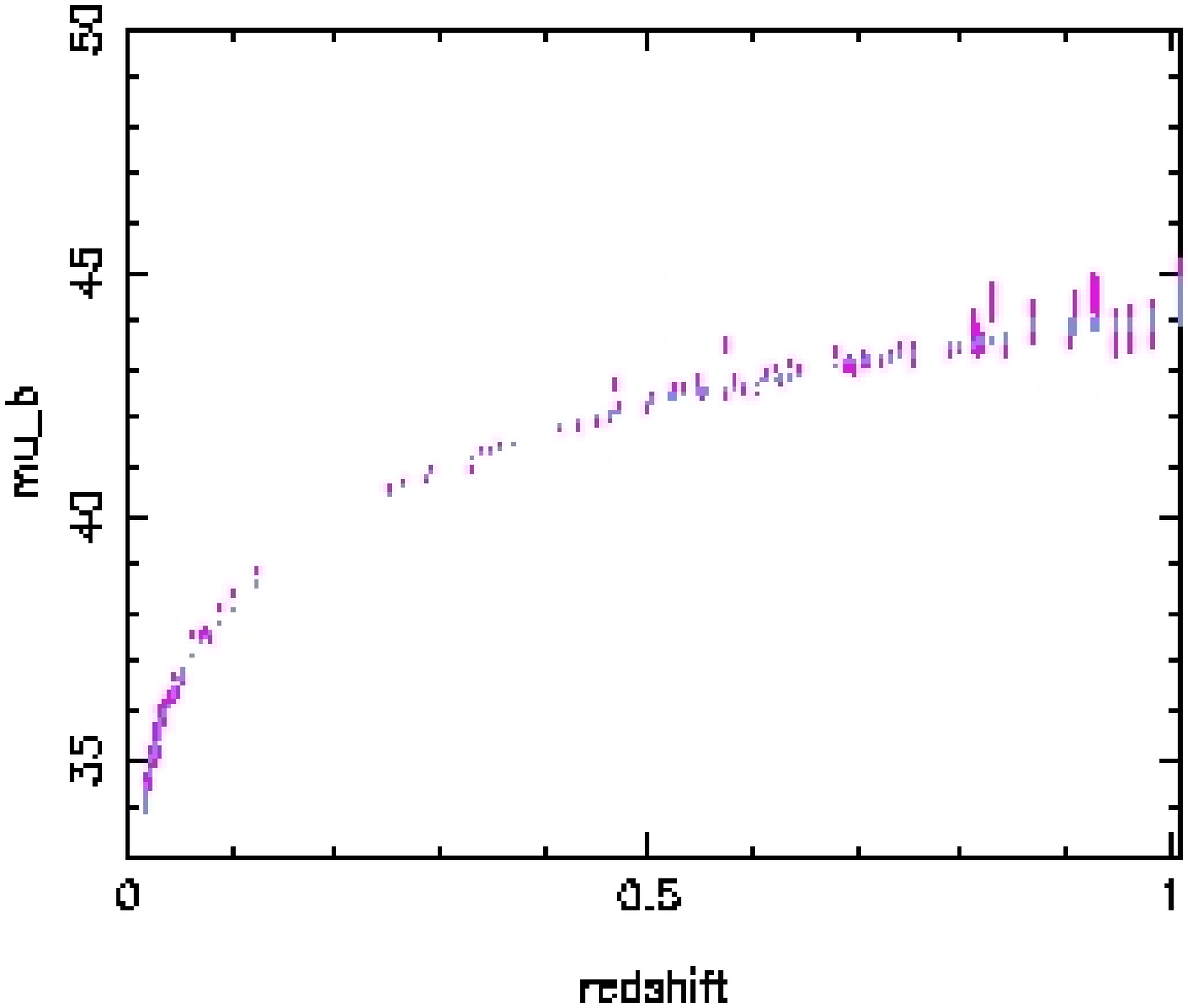} &
{\bf b} \includegraphics[width=6cm]{\filepath/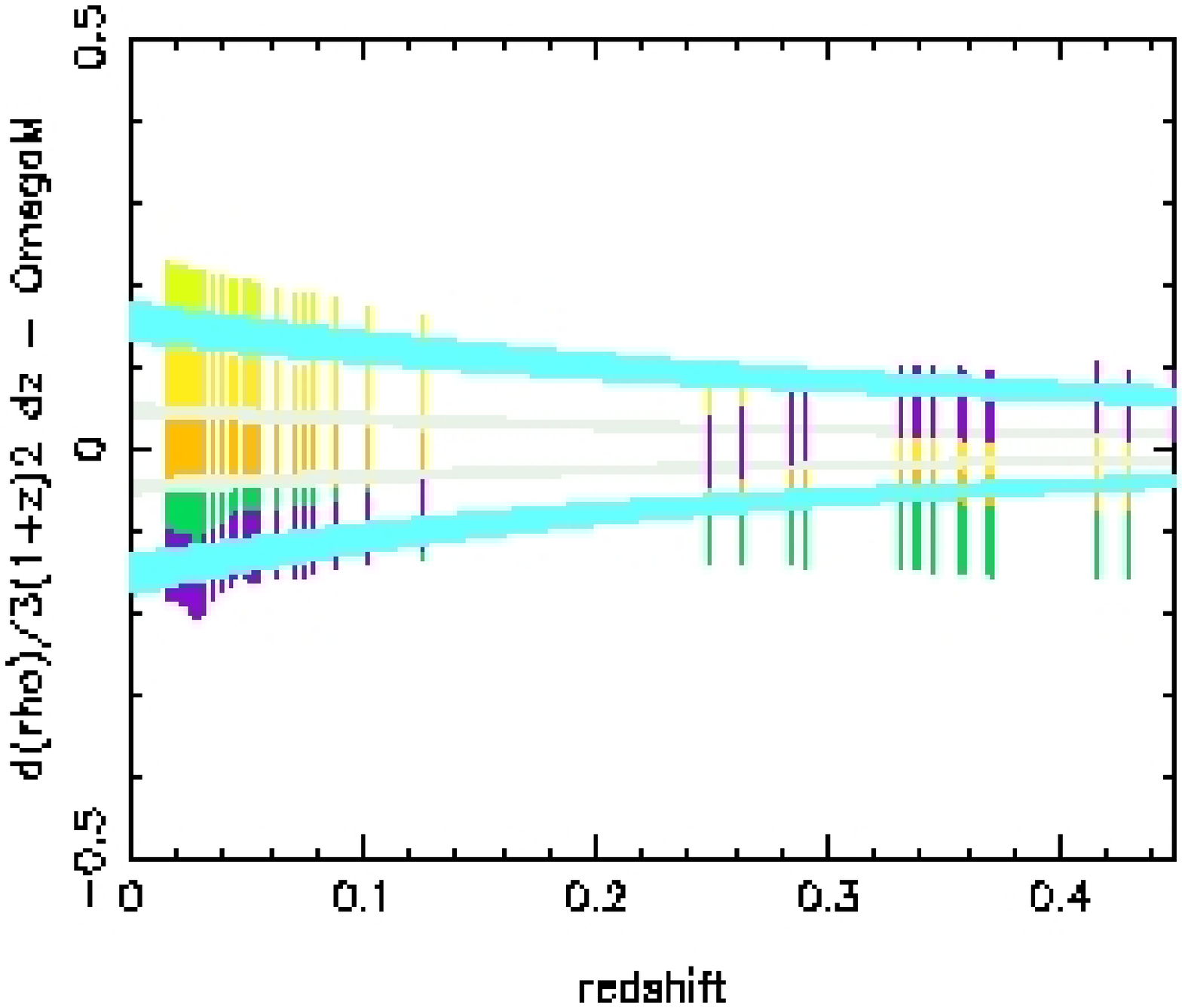}
\end{tabular}
\caption{{\bf a}: $\mu_b$ from SNLS supernovae, data (magenta), smoothed 
distribution (blue). Although this distribution look quite smooth, even small 
sudden variations can lead to large variations in derivatives. {\bf b}: 
${\mathcal A}(z)$ for SNLS supernovae with $z < 0.45$. Definition of curves 
is the same as in Fig.\ref{fig:sncp}. For this dataset $1-\bar{P} = 0.93$ when 
cosmological parameters are the same as Fig.\ref{fig:sncp}. 
If $\Omega_{de} = 0.73$ is used, $1-\bar{P} = 0.96$.
\label{fig:snlslow}}
\end{center}
\end{figure}

We have applied this algorithm to two supernova datasets: 
published data from SuperNova Legacy Survey (SNLS)\cite{snls}, and to 
supernovae with $z < 0.45$ from gold sample of re-analyzed data by 
Riess \etal 2004\cite{sncp}. The reason for using only the low 
redshift subset of the latter compilation is that scatter and uncertainties 
of the peak magnitude at higher redshifts is too large and reconstructed 
${\mathcal A}(z)$ is instable. Fig.\ref{fig:sncp} shows ${\mathcal A}(z)$ 
obtained from these data as well as from simulated SNe as described in the 
caption. Simulated standard sources are put at the same redshifts as real 
data to make simulated samples as similar 
to data as possible. In both datasets the probability of $\gamma \lesssim 0$ 
or equivalently $w \lesssim -1$ is larger than $70\%$. The SNLS data is 
consistent with a $\gamma$ as small as $\sim -0.2$\footnote{Here we 
concentrate on the sign of $\gamma$ and don't perform any fitting to obtain 
its value. Estimations are from comparison with simulations.}. There is 
however significant deviation from a smooth distribution for $z \lesssim 0.1$ 
and $z \gtrsim 0.5$. We attribute this to large scatter of the data at these 
redshifts see Fig.\ref{fig:snlslow}-{\bf a}. The fact that at intermediate 
redshifts in these datasets $A(z)$ follows closely models with constant 
$\gamma$ is a demonstration of very small variation of $\gamma$ in these 
redshifts, consistent with our arguments above.

To see whether the large negative $\gamma$ concluded from the SNLS data is due 
to scattering and/or reconstruction artifacts, we also apply the same 
formalism to a subset of these data with $z < 0.45$. The result is shown in 
Fig.\ref{fig:snlslow}-{\bf b} along with simulations in the same 
redshift range. The {\it bump} at very low redshifts in Fig.\ref{fig:sncp} 
does not exist in this plot, and therefore we conclude that it has been an 
artifact of numerical reconstruction. Although ${\mathcal A}(z)$ distribution 
in this dataset is also convex and the probability of $\gamma < 0$ is larger 
than $90\%$, it does not have the same slope as any of constant $w$ models. 
It seems that low and high redshift sections of the curve correspond to 
different values of $\gamma$. For $z \lesssim 0.15$, ${\mathcal A}(z)$ is 
close to theoretical and simulated data with $\gamma = -0.2$. For 
$z \gtrsim 0.25$, 
${\mathcal A}(z)$ approaches the values for larger and even positive $\gamma$. 
Such behaviour does not appear in Fig.\ref{fig:sncp}. Giving the fact that the 
number of available data points with $z \gtrsim 0.25$ in this subset is small, 
the most plausible explanation is simply numerical artifacts. Alternative 
explanations are the evolution of $\gamma$ with redshift or the use of 
under-estimated value for $\Omega_m$. If the latter case is ture, the value 
of $\gamma$ must be even smaller than $-0.2$. Interestingly, the deviation 
from a constant $w$ model in the former case is consistent, up to 
uncertainties, with the best estimations of the evolution of $w$ in models 
IV and VI of Ref.~\cite{wvar}. This confirms the consistency of two 
reconstruction and parameter estimation methods. However, as explain above, 
present method shows that this deviation can be also due to uncertainties of 
$\Omega_m$ and not the evolution of $w$. The simplicity of dependence on the  
cosmological parameters in this method permits to see their effects more 
explicitly than in fits. With a data gap in $0.15 \lesssim z \lesssim 0.25$ 
interval and only 58 supernovae in this subset, it is not possible to make 
any definite conclusion about the behaviour of this data. We should also 
mention that SNLS supernovae with $z < 0.25$, and at $0.25 < z < 0.4$, and 
$ z > 0.4$ have not been treated in the same way\cite{snls}. It is therefore 
possible that some of the observed features are purely artifacts of the 
analysis of the raw observations. 

Fig.\ref{fig:sncp}-{\bf b} shows ${\mathcal A}(z)$ from gold 
SNe sample recompiled by Riess, \etal\cite{sncp}. It is also consistent with 
$\gamma < 0$ with a probability $\sim 75\%$ at 1-sigma and $\sim 66\%$ at 
2-sigma. This plot shows that the value of $\gamma$ estimated from this 
data is $\sim -0.06$, larger than estimation from SNLS data.

The reason for the difference between estimated values for $\gamma$ from SNLS 
and Riess, \etal compilation here is not clear because value of $w$ 
reported in Ref.\cite{sncp} and Ref\cite{snls} are consistent. Nonetheless, 
the estimated values are in the range reported by other 
works\cite{wmapeyear}\cite{cmagicsn}. Most probably this difference is 
related to different scatter and uncertainty of these datasets, and the fact 
that low and high redshift data are not treated in the same way. Recent 
claims about contamination of supernova type Ic and the effect of asymmetric 
explosion in the lightcurve of supernova type Ia\cite{iaic}, and possible 
differences between low and high redshift supernovae\cite{snlowhigh} can not 
explain the differences either, because they affect both surveys in 
the same way. Despite these discrepancies, both datasets are in good agreement 
about negative sign of $\gamma$.

These results highlight various shortcomings in both datasets used here. Our 
first remark is the large gap in redshift distribution of 
observed supernovae in redshift range $0.1 \lesssim z \lesssim 0.3$. Both 
datasets have less than 6 supernovae in this range and nothing in 
$0.15 \lesssim z \lesssim 0.25$. This is not an important issue for finding 
redshift distribution of $D_l$, but redshift gap becomes very important when 
derivatives of $D_l$ are calculated. The lack or rareness of supernovae data 
in this redshift range is related partly to the history of star formation 
in galaxies\cite{sfhist}, and partly to optimization of 
surveys\cite{snzdist} for detection of very low or high redshift supernovae 
which decreases the probability of detection of mid-range SNe. Sloan Supernova 
Survey\cite{sdsssn} is optimized to detect SNe in 
$0.1 \lesssim z \lesssim 0.35$ and should provide the missing data in near 
future. 

Our second observation is a large scatter in both datasets around 
redshift $\sim 0.55$, see Fig.\ref{fig:snlslow}-{\bf a}. This leads to a 
large scatter in the numerical determination of derivatives in 
(\ref {rhoderdl}) and makes the results unusable. In future observations 
the reason of this large scatter should be understood. 

From theoretical calculation and simulations shown in Fig.\ref{fig:sncp} one 
can also conclude that with present uncertainties of 
cosmological parameters, the most important redshift range for determining 
the equation of state of dark energy is $z \lesssim 0.8$. Higher redshift 
supernovae are only interesting if $w$ varies significantly with redshift. 
Although technical challenges, understanding of physics of supernovae 
and their evolution\cite{snlowhigh}, and applications for other astronomical 
ends make the search of supernovae at larger distances interesting, they will 
not be in much use for determining the equation of state of dark energy, at 
least not at the lowest level which is the determination of redshift 
independent component of $w$. 

On the other hand, improvements in numerical techniques and algorithms 
would lead to better measurements. One of the possibilities in this direction 
is the application of an adaptive smoothing algorithm with variable degrees of 
smoothing depending on the amount of scatter in the data. More 
sophisticated smoothing algorithms also have been proposed\cite{snsmooth}. We 
postpone the application of these advanced methods to future, when larger 
datasets become available.

In summary, we have proposed a nonparametric formalism to investigate the 
sign of $\gamma$ in the equation of state of dark energy. When data with 
better quality become available, fitting can be added to this method to find 
also the value of $\gamma$. The advantage of this method is its geometrical 
nature when the sign of $\gamma$ is searched. Moreover, its simple dependence 
on cosmological parameters makes the assessment of uncertainties of the value 
of $\gamma$ easier. This method is suitable for applying to observations in 
which the total density of the Universe and its variation are directly 
measured such as standard candles and galaxy surveys, but not to integrated 
observables such as CMB anisotropy. By applying this method to two of largest 
publicly available supernovae datasets we showed that they are consistent 
with a $w < -1$. Present data is not however enough precise to permit the 
estimation of $|w+1|$ with good certainty. With on-going projects such as 
SNLS, Supernova Cosmology Project\cite {realsncp}, and SDSS SNe survey, and 
future projects such as SNAP and DUNE, enough precise datasets should be 
available soon.

{\bf Appendix:} For standard candles, ${\mathcal A}(z)$ must be calculated 
from luminosity distance which is related to magnitude: 
$D_l/D_0 = 10^{\mu_b / 5}$. $D_0$ is the distance for which the common 
luminosity of standards is determined and depends on $H_0$. Therefore, 
$\mu_b$ needs a correction if a different $H_0$ is used\cite{snls}. For 
simulated data, $D_l$ is calculated from (\ref{dlh}) and $\mu_b$ is 
determined from definition above, then a random noise with a standard 
deviation of 3\% is added to the magnitude. 

Expression (\ref{rhoderdl}) for ${\mathcal B}(z)$ contains the first and 
the second derivatives of $D_l$ which must be calculated numerically from 
data. It is however well known that a direct determination of derivatives 
leads to large and unacceptable deviation from exact values. One of the most 
popular alternatives is fitting a polynomial around each data point and 
then calculating an analytical derivative using the polynomial approximation 
in place of the data. We use this approach to determine derivatives of $\mu_b$ 
and $D_l$. In addition, before applying this approach, we smooth the 
distribution of magnitudes using again the same polynomial fitting algorithm. 
In FLRW cosmologies the redshift evolution of the luminosity distance is 
very smooth. Therefore, a second order polynomial for smoothing is adequate.
Fitting is based on a right-left symmetric, least $\chi^2$ algorithm, and 
for this we have implemented {\it lfit} function of Numerical 
Recipes\cite{numrec}. By trial and error we find that 19 data-point fitting 
gives the most optimal results regarding the number of available data points 
and their scatter. Close to boundaries however less data point for fitting is 
available in one side of each point, and therefore the fitting is less 
precise. The artifacts discussed above are mostly related to this imprecision 
of calculation near boundaries. In the present work no adaptive smoothing is 
applied. In addition to smoothed data and their derivatives, the function 
{\it lfit} calculates a covariant matrix for uncertainties of parameters 
(derivatives). We use diagonal elements as 1-sigma uncertainty of the smoothed 
data and its derivatives. The uncertainty of ${\mathcal A}(z)$ is calculated 
from the uncertainty of terms in (\ref{densderiv}) and (\ref{rhoderdl}) using 
error propagation relation: For $f(x_1, x_2, \ldots)$, $\sigma_f^2 = \sum_i 
\sigma^2_{x_i} (\partial f/\partial x_i)^2$. Smoothed terms and parameters 
in ${\mathcal A}(z)$ are considered as independent variables with their own 
uncertainty.

\end{document}